\documentclass[a4,11pt]{article}
\usepackage{latexsym,enumerate}
\usepackage{amsmath,amsthm,amsopn,amstext,amscd,amsfonts,amssymb}
\usepackage{natbib,graphicx,fullpage}
\usepackage{hyperref}
\usepackage{amssymb}
\usepackage{graphicx}
\usepackage{bbm}

\begin{document}

\noindent
{\Large \bf Discussion on ``Random-projection ensemble classification" by T.~Cannings and R.~Samworth, written by Roberto Casarin$^\dag$, Lorenzo Frattarolo$^\dag$ and Luca Rossini$^{\dag\ddag}$\footnote[1]{Corresponding author at: Free University of Bozen-Bolzano, piazza Universit\`a, 1, 39100, Bolzano-Bozen, Italy. \\ \textit{E-mail address}: \href{luca.rossini@unive.it}{luca.rossini@unive.it} (Luca Rossini)}({\large $^\dag$University Ca' Foscari of Venice, Italy and $^\ddag$ Free University of Bozen-Bolzano, Italy}).}

\vspace{10pt}

The authors are to be congratulated on their excellent research, which has culminated in the development of a characterization of the  approximation errors in random projection methods when applied to classification. We believe that the proposed approach can find many applications in economics such as credit scoring (e.g. \cite{Altam68}) and can be extended to more general type of classifiers. In this discussion we would like to draw authors attention to the copula-based discriminant analysis (\cite{CODA13} and \cite{HeZhangWang16}).

We consider $X|Y=r$ distributed as a $p$-dimensional meta Gaussian distribution and $\left.S\right\vert Y=r\sim \mathcal{N}_p\left(0,\Sigma_r\right)$, where $\Sigma_r$ is the linear correlation among variables. Given a $p \times d$ random projection $A$, $\left.AS\right\vert Y=r\sim \mathcal{N}_d\left(0,\Sigma^{A}_r\right)$, where $\Sigma^{A}_r= A\Sigma_r A^{T}$. If we assume that the information in the marginals is not relevant for the classification, the Bayes decision boundary depends only on the transformed variables $s_i=\Phi^{-1}\left(F(x_i)\right)$ with $\Phi$ and $F$ the univariate normal and the marginal cdfs, respectively (\cite{FANG20021}), $s_i$ and $x_i$ the $i$-th element of $s$ and $x$, and the correlation of the two groups
 \begin{eqnarray}
 \Delta\left(s; \pi_0, \Sigma_0,\Sigma_1\right) = \log\left(\dfrac{\pi_1}{\pi_0}\right) - \dfrac{1}{2} \log\left(\dfrac{\det\left(\Sigma_1\right)}{\det\left(\Sigma_0\right)}\right)- \dfrac{1}{2} s^{T}\left[ \Sigma_1^{-1}-\Sigma_0^{-1}\right]s.
 \end{eqnarray} 
Analogously the classifier in the random projection ensamble will depend only on the random projection of the transformed variables and their covariances. We use the empirical distribution function to obtain the sample version of the transformed variables $S_i=(S_{1i},\dots,S_{pi})$, with
\begin{eqnarray}
S_{ji}&=&\Phi^{-1}\left(\dfrac{1}{n+1}\sum^n_{k=1} \mathbbm{1}_{\left\lbrace X_{jk}\leq X_{ji}\right\rbrace}\right), \quad
 i=1,\ldots,n, \quad j=1,\ldots,p.
\end{eqnarray}
The estimator of $\Sigma^{A}_r$ is obtained by maximizing the pseudo-likelihood:
\begin{eqnarray*}
\hat{\Sigma}^{A}_r&=& \frac{1}{n}\sum^{n}_{i=1} A S_i S_i^{T}A^{T}\mathbbm{1}_{\left\lbrace Y_i^{A}=r\right\rbrace} \quad \mbox{ for } r=0,1
\end{eqnarray*}
where the asymptotic normality is guaranteed by results in \cite{Genest1995} and recently in \cite{segers2014}. We propose the following robust QDA random-projection ensemble classifier:
\begin{eqnarray}
C_{n}^{\text{A-RQDA}}(s):=\left\lbrace\begin{array}{ccc}
1 &&\Delta\left(s; \hat{\pi}_0, \hat{\Sigma}^{A}_0,\hat{\Sigma}^{A}_1\right)\geq 0, \\ 0 && \mathrm{otherwise}.
\end{array}\right.
\end{eqnarray}
We are very pleased to be able to propose the vote of thanks to the authors for their work.

\bibliographystyle{apalike}
\bibliography{Refer}

\end{document}